\documentclass[12pt]{amsart}
\usepackage{graphics}
\usepackage{amsthm}
\usepackage{amssymb}
\usepackage{amsmath}
\usepackage{amsfonts}
\usepackage{epic}
\usepackage{curves}
\usepackage{epsfig}
\usepackage{bm}
\usepackage{amscd}
\input{psfig.sty}

\newcommand{\baseRing}[1]{\ensuremath{\mathbb{#1}}}
\newcommand{\Z}{\baseRing{Z}}
\newcommand{\C}{\baseRing{C}}

\newcommand{\R}{\baseRing{R}}

\theoremstyle{plain}

\theoremstyle{definition}

\def\be{\begin{equation}}
\def\ee{\end{equation}} 
\title{On Spectral Flow Symmetry and Knizhnik-Zamolodchikov Equation}

\author{Gast\'on E. Giribet}
\address{Departamento de F\'{\i}sica de la Universidad de Buenos Aires e Instituto de F\'{\i}sica de la Universidad Nacional de La Plata.}
\thanks{Based on a talk given at the Workshop on Non-Perturbative Gauge Dynamics (SISSA, Trieste, 2005.)}

\begin{document}

%%%%%%%%%%%%%%%%%%%%%%%%%%%%%%%%%%%%%%%%%%%%%%%%%%%%%%%%%%%%%%%%%%%%%%%%

\begin{abstract}

It is well known that five-point function in Liouville field theory provides a representation of
solutions of the $SL(2,\R )_k$ Knizhnik-Zamolodchikov equation at the level of
four-point function. Here, we make use of such representation to study some aspects
of the spectral flow symmetry of $\hat {sl(2)}_k$ affine algebra and its
action on the observables of the WZNW theory. To illustrate the usefulness of
this method we rederive the
three-point function that violates the winding number in $SL(2,\R )$ in a very succinct way. In addition,
we prove several identities holding between exact solutions of the
Knizhnik-Zamolodchikov equation.

\end{abstract}

%%%%%%%%%%%%%%%%%%%%%%%%%%%%%%%%%%%%%%%%%%%%%%%%%%%%%%%%%%%%%%%%%%%%%%%%

\maketitle

\section{Introduction}

Spectral flow symmetry of $\hat {sl(2)}_k$ affine algebra induces
the
identification between some states belonging to discrete representations of
$SL(2,\R ) _k$; namely the identification between states of the of representations ${\mathcal D}^{\pm }_{j}$ and ${\mathcal D}^{\mp }_{k/2 -j}$ (see Ref. \cite{MaldacenaOoguri1}; see also Ref. \cite{Halpern} for previous studies on spectral flow symmetry). At the level of four-point correlation functions, such identification
is realized by identities between different solutions of the Knizhnik-Zamolodchikov equation
(KZ) (see Ref. \cite{NicholsSanjay}). In Ref. \cite{GiribetSimeone4} it was pointed out that the description of the four-point functions of the $SL(2,\R )_k$ CFT in terms of the Liouville five-point function leads one to interpret
the action of the spectral flow as yielding a simple Liouville reflection
transformation. Hence, this ``Liouville description'' of the $SL(2,\R )_k$ correlators
turns
out to be a useful tool to work out the details of certain symmetries of KZ
equation which, otherwise, would remain hidden within the original picture.

Actually, the aim of this note is that of studying the manifestation of the spectral flow symmetry as encoded in the symmetries of the four-point correlators by means of this method. Moreover, we extend the discussion to the case of other $\Z _2$ transformations.

\subsection{The method}

\subsubsection*{The five point function in Liouville field theory}
Let us begin by considering the quantum Liouville field theory of central charge $c=1+6Q^2$, with $Q=b+b^{-1}$. The Liouville field
$\varphi (z)$ transforms under holomorphic coordinate transformations $z\to z'$ as
follows
\begin{equation}
\varphi (z) \to \varphi (z') = \varphi (z)+Q \log \left| \frac {dz}{dz'}\right|
\end{equation}
and satisfies the asymptotic behavior $\varphi (z)+2Q\log |z|\sim {\mathcal O}(1)$ for large $|z|$. This specifies the
boundary conditions for the theory on the sphere (see \cite{Nakayama} for an excellent review). Here, we want to consider correlation functions involving local primary
operators $V_{\alpha }(z)$ with conformal dimension $\Delta _{\alpha }= \alpha (Q-\alpha )$. These operators can
be realized by the exponential form $V_{\alpha }(z)\sim e^{\sqrt {2}\varphi (z)}$, where the Liouville field that satisfy the free field correlator $<\varphi (z) \varphi (0)>=-2\log |z|$. These obey
\begin{equation}
V_{Q-\alpha } (z) = R_b (\alpha ) V_{\alpha } (z) \ ; \ \ \langle V_{\alpha }(z_1) V_{\alpha }(z_2)\rangle = |z_1-z_2|^{-2\alpha (Q-\alpha )} R_b (\alpha )   \label{reflection}
\end{equation}
where $R_b (\alpha )$ is the Liouville reflection coefficient, given by
\begin{equation}
R_b (\alpha ) = - \left( \pi \mu b^{-2b^2} \frac {\Gamma (1+b^2)}{\Gamma (1-b^{2})} \right)^{1+b^{-2}-2\alpha b^{-1}}b^{-2} \frac {\Gamma (2b\alpha -b^2)\Gamma (1+b^{-2}-2b^{-1}\alpha )}{\Gamma (2b^{-1}\alpha -b^{-2})\Gamma (1+b^2-2b\alpha )}
\end{equation}
and where the Liouville cosmological constant $\mu $ can be choosen as $\mu = \pi ^{-2}$ for convenience. A consequence of (\ref{reflection}) is the following operator valued relation
\begin{equation}
\langle  V_{\alpha _1}(z_1)  V_{\alpha _2}(z_2) ... V_{\alpha _N}(z_N) \rangle = R_b (\alpha _1)\langle  V_{Q-\alpha _1}(z_1)  V_{\alpha _2}(z_2) ... V_{\alpha _N}(z_N) \rangle  \label{reflex}
\end{equation}
and the same for any of the $N$ vertex operators.

For our purpose, we are interested in the five-point function ($N=5$). This will be denoted as
\begin{eqnarray}
{\mathcal A}^{Liouville}_{\alpha _1,\alpha 
_2,\alpha _5,\alpha_ 3,\alpha _4} (x,z) = \langle V_{\alpha _2} (0) V_{\alpha _1} (z) V_{\alpha _5} (x) V_{\alpha _3} (1) V_{\alpha _4} (\infty ) \rangle .   \label{5p}
\end{eqnarray}
The conformal blocks of this correlators and the connection with the analogous quantities in WZNW theory was discussed in detail in \cite{Ponsot}.

Now, let us move to the WZNW theory which is the second ingredient for the method to be used.

\subsubsection*{The four-point functions in the $SL(2,\R )_k$ WZNW model}

We consider the $SL(2,\R )_k$ WZNW model. In particular, we concern about the four-point function. This observable is constructed by a sum over the conformal blocks of the theory. Conformal blocks are solutions of the Knizhnik-Zamolodchikov equation labeled by the internal quantum number $j$ that is interpreted as the momentum of the interchanged states in the factorization procedure. Here, we are interested in correlators of the form
\begin{eqnarray}
{\mathcal A}^{WZNW}_{j_1,j_2,j_3,j_4} (x,z) = \langle \Phi _{j_2}(0,0) \Phi _{j_1}(x,z) \Phi _{j_3}(1,1) \Phi _{j_4}(\infty ,\infty ) \rangle  \label{4p}
\end{eqnarray}
that involves the local primary fields $\Phi _j (x,z)$. These operators are associated to the analytic continuation of
differentiable functions on the $H_3^+=SL(2,\C )/SU(2)$. The complex variable $x$ permits to order the representations of
$SL(2,\R )$ as follows
\begin{equation}
J^a (z) \Phi _j(x,z')= \frac {1}{(z-z')} {D}^a_x \Phi _j (x,z) + ...
\end{equation}
where $a \in \{ \pm , 3 \} $ correspond to the $SL(2,\R )$ generators $J ^a$, realized by
\begin{equation}
 D_x^- = x^2 \partial _x -2jx \ , \ \ D_x^+ = \partial _x \ , \ \ D_x^3 = x\partial _x -j
\end{equation}
This representation is often employed in the applications to string theory in $AdS_3$ space since it provides a clear interpretation of the $AdS_3 /CFT_2$ correspondence (see \cite{KutasovSeiberg} for a detailed description of functions $\Phi _j (x,z)$ within this context).

\subsubsection*{The Fateev-Zamolodchikov correspondence}

Once correlation functions in both Liouville and WZNW theories were introduced, we undertake the task of connecting them. Certainly, it is feasible to express the four-point functions ${\mathcal 
A}^{WZNW}_{j_1,j_2,j_3,j_4}$ in terms of the five-point functions ${\mathcal A}^{Liouville}_{\alpha _1,\alpha _2,-\frac{1}{2b},\alpha_ 
3,\alpha _4}$ (including a particular fifth state $\alpha _5 
=-\frac{1}{2b}$); namely
\begin{eqnarray}
{\mathcal A}^{WZNW}_{j_1,j_2,j_3,j_4} (x,z) &=& X_k (j_1,j_2,j_3,j_4|x,z) F_k (j_1,j_2,j_3,j_4){\mathcal A}^{Liouville}_{\alpha _1,\alpha 
_2,-\frac{1}{2b},\alpha_ 3,\alpha _4} (x,z) \label{fz}
\end{eqnarray}
where $c_b$ represents a $j_{\mu }$-independent numerical factor whose explicit form can be found in the literature (see \cite{FateevZamolodchikov,Andreev,Teschner,Ponsot,GiribetSimeone4}); besides, the quantum numbers of both correlators are related by
\begin{eqnarray}
2\alpha _{1} =b\left( j_{1}+j_{2}+j_{3}+j_{4}-1\right) , \ \ 2\alpha _{i} =b\left( j_{1}-j_{2}-j_{3}-j_{4}+2j_{i}+b^{-2}+1\right)  \label{mapa}
\end{eqnarray}
being $i \in \{ 2,3,4 \}$, and
\begin{eqnarray}
 2\alpha _5 =-b^{-1} \ ,\ \  b^{-2}=k-2 \label{mapk}
\end{eqnarray}
The normalization factors are given by
\begin{eqnarray}
X_k(j_1,j_2,j_3,j_4|x,z) = |x|^{-2\alpha _{2}/b}|1-x|^{-2\alpha
_{3}/b}|x-z|^{-2\alpha _{1}/b}|z|^{-4(b^{2}j_{1}j_{2}-\alpha _{1}\alpha
_{2})}|1-z|^{-4(b^{2}j_{3}j_{1}-\alpha _{3}\alpha _{1})} \nonumber
\end{eqnarray}
and
\begin{eqnarray}
F_k (j_1,j_2,j_3,j_4) &=& c_b \left( \pi b^{2b^2} \frac {\Gamma (1-b^2)}{\Gamma (1+b^2)}\right) ^{1+j_1-j_2-j_3- j_4} \ \prod _{\mu = 1}^4 \frac {\Upsilon _b (2j_{\mu}b-b)}{\Upsilon _b (2\alpha _{\mu})} \label{F}
\end{eqnarray}
where the $\Upsilon _b$ function is defined as follows
\[
\log \Upsilon _b (x)=\frac{1}{4}\int_{\R _{>}}\frac{d\tau }{\tau }\left(
Q-2x\right) ^{2}e^{-\tau }-\int_{\R _>}\frac{d\tau }{\tau }%
\frac{\sinh ^{2}\left( \frac{\tau }{4}(Q-2x)\right) }{\sinh \left( 
\frac{b\tau }{2}\right) \sinh \left( \frac{\tau }{2b}\right) } 
\]
This function presents the zeros at $x \in -b\Z_{\geq 0}-b^{-1}\Z_{\geq 0}$ and $ x \in b\Z_{>0}+b^{-1}\Z_{>0}$

\subsubsection*{Remarks}

Normalization factor (\ref{F}) is the appropriate to connect the correlators of both models, leading to the correct structure constants when one of the momenta tends to zero. Besides, the scaling $\left(\pi b^{2b^2}\frac {\Gamma (1-b^2)}{\Gamma (1+b^2)}\right)^{1+j_1-j_2-j_3-j_4} $ corresponds to having set the Liouville cosmological constant to a specific value, namely $\mu = \pi ^{-2}$. Such factor is not symmetric under permutations of the symbol $\{j_1,j_2,j_3,j_4\}$, as the map (\ref{mapa}) is not; this is to make the KPZ scaling of both correlators to match.

We can also understand the presence of the fifth vertex at $x$ with momentum $\alpha _5=-\frac {1}{2b}$. This is certainly related to the existence of degenerate representations of $SL(2,\R )_k$ ({\it i.e.} those representations containing null states in the modulo). Some of these representations are those having momentum $j$ such that $1-2j=m\in \Z _{>0}$. According to the conformal Ward identities, the correlators involving an operator $\Phi _{\frac {1-m}{2}}(x,z)$ are annihilated by the differential operator $\partial _x^m$ (and similarly for $\partial _{\bar x}^m$). This corresponds to the fact that, if $1-2j_1$ is a positive integer, then $\Phi _{j_1}(x,z)$ turns out to be a polynomial of degree $m-1$. Besides, (\ref{mapa}) implies that, when realizing the Liouville correlator (\ref{5p}) in terms of the Coulomb gas-like prescription, the amount of screening charges to be employed is exactly $n=-2j=m-1$. Hence, a simple computation leads to obtain
\begin{eqnarray}
(D_x^+)^m X_k ( (1-m)/{2}, j_2,j_3,j_4|x,z) {\mathcal A}^{Liouville}_{\alpha _1, \alpha _2, -\frac {1}{2b},\alpha _3,\alpha _4} (x,z) = |z|^{-4b^2j_1j_2} |1-z|^{-4b^2j_1j_3} \times \nonumber \\ 
\times \partial _x^m \prod _{r=1}^{m-1} \int d^2 w_r |w_r|^{-4\alpha _2 b}|1-w_r|^{-4\alpha _3 b} |z-w_r|^{-4\alpha _1 b}|x-w_r|^{2} =0
\end{eqnarray}
which is immediately obeyed due to the fact that the integrand is a polynomial of degree $m-1$ in $x$ (and the same for $\bar x$); and this is a direct consequence of the OPE $e^{\sqrt {2} \alpha _5 \varphi (x)} e^{\sqrt {2} b \varphi (w_r)} \sim |x-w_r|^{2}+...$ which holds for $\alpha _5 = -\frac {1}{2b}$.

The Liouville description of WZNW correlators was successfully employed in
working out several details of the non-compact $SL(2,\C )_k/SU(2)$ CFT. For
instance, the crossing symmetry of the theory was proven in \cite{Teschner}
and some aspects of the singularities in the WZNW observables were
understood by means of this method (for instance, see \cite{Ponsot} and \cite{GiribetSimeone4}). Conversely, the act of thinking the
four-point correlators (\ref{4p}) as a five-point function of other CFT
({\it i.e.} the Liouville CFT) permitted to understand the arising of
certain poles at the middle of the moduli space; namely at $z=x$. In terms
of the function (\ref{5p}) these poles are understood as coming from the factorization
limit when the operators $V_{\alpha _1}(z)$ and $V_{\alpha _5}(x)$ coincide.

\subsection{Outline}

We will study solutions of the Knizhnik-Zamolodchikov equation by means of their connection with correlators in Liouville theory. The paper is organized as follows: In the next section we analyze the action of the spectral flow symmetry on the four-point correlation function in the WZNW model. We explicitly show how the identification between states of discrete representations ${\mathcal D}_{j}^{\pm }$ and ${\mathcal D}_{k/2-j}^{\mp }$ of $SL(2,\R )_k$ turns out to correspond to the Liouville reflection of one particular vertex operator. This was pointed out in \cite{GiribetSimeone4}. This simple observation leads us to rederive the formula for the three-point violating winding amplitude in $AdS_3$. The key point in doing this is the normalization (\ref{F}), which encodes the information of the WZNW structure constants. 

In section 3 we apply the method of describing WZNW correlators in terms of their Liouville analogues to study other symmetries of KZ equation. Then, we are able to prove some identities between exact solutions in a rather simple way. This permits to visualize hidden symmetries of the KZ equation turning them ``expectable''.

\section{Spectral flow symmetry and the Knizhnik-Zamolodchikov equation}

\subsection{The Three-Point Function Violating the Winding Number}

First, let us consider the following 4-point correlation function in the $SL(2,\R )_k$ WZNW model;
\begin{equation}
{\mathcal A}^{WZNW}_{\frac k2,j,j_3,j_4} (x,z) =  \langle \Phi _{j} (0,0) \Phi _{\frac k2} (x,z) \Phi _{j_3} (1,1)  \Phi _{j_4} (\infty ,\infty )  \rangle
\end{equation}
This particular quantity represents the three-string scattering amplitude in $AdS_3$ space for the case of non-conserved total winding number. This was computed in \cite{MaldacenaOoguri3}, and the calculation employs the inclusion of the additional (auxiliary) vertex $\Phi _{\frac k2}(x,z)$ which
is often called ``spectral flow operator'' or ``conjugate representation of the
identity operator''.

%%%%%%%%%%%%%%%%%%%%%%%%%%%%%%%%%%%%%%%%%%%%%%%%%%%%%%

By using the correspondence (\ref{fz}) and (\ref{mapa}), we find the following expression
\begin{eqnarray}
\langle \Phi _{j} (0,0) \Phi _{\frac k2} (x,z) \Phi _{j_3} (1,1)  \Phi _{j_4} (\infty ,\infty )  \rangle =  X_k (k/2,j,j_3,j_4|x,z) F_k (k/2,j,j_3,j_4) \times \nonumber \\ \times \prod _{\nu =1}^{4} R_b(\alpha _{\nu }) \langle V_{Q-\alpha _2}(0)  V_{Q-\alpha _1}(z) V_{-\frac {1}{2b}}(x) V_{Q-\alpha _3}(1) V_{Q-\alpha _4}(\infty )  \rangle \label{otra}
\end{eqnarray}
where $2\alpha _1 = b^{-1}(j+j_3+j_4+b^{-2}/2)$, $2\alpha _2 = b^{-1}(j-j_3-j_4+3b^{-2}/2+2)$, $2\alpha _3 = b^{-1}(-j+j_3-j_4+3b^{-2}/2+2)$ and  $2\alpha _4 = b^{-1}(-j-j_3+j_4+3b^{-2}/2+2)$. A crucial observation is that, according to (\ref{mapa}), the condition
$j_1=k/2$ implies the constraint $\sum _{\mu =1}^4 \alpha _{\mu }=\frac 52
b^{-1}+ 3b$. Hence, the corresponding Liouville five-point function can be realized by using a Coulomb-gas like prescription with no insertion of screening charges; this is due to the fact that the identity $\sum _{\nu =1}^4 (Q-\alpha _{\nu })+nb -\frac {1}{2b}=Q$ is obeyed precisely for $n=0$. Then, we have the realization
\begin{eqnarray}
\langle e^{\sqrt {2}(Q-\alpha _1) \varphi (z)} e^{\sqrt {2}(Q-\alpha _2) \varphi (0)} e^{\sqrt {2}(Q-\alpha _3) \varphi (1)} e^{-\frac {1}{\sqrt {2}b} \varphi (x)} e^{\sqrt {2}(Q-\alpha _4) \varphi (\infty )} \rangle = \nonumber \\ |z|^{4(Q-\alpha _1)(\alpha _2-Q)}|1-z|^{4(Q-\alpha _1)(\alpha _3-Q)}|x-z|^{2b^{-1}(Q-\alpha _1)}|x|^{2b^{-1}(Q-\alpha _2)}|1-x|^{2b^{-1}(Q-\alpha _3)} \nonumber
\end{eqnarray}
and, from (\ref{mapa}), we eventually find
\begin{eqnarray}
X_k (k/2,j,j_3,j_4|x,z)\langle e^{\sqrt {2}(Q-\alpha _1) \varphi (z)} e^{\sqrt {2}(Q-\alpha _2) \varphi (0)} e^{\sqrt {2}(Q-\alpha _3) \varphi (1)} e^{-\frac {1}{\sqrt {2}b} \varphi (x)} e^{\sqrt {2}(Q-\alpha _4) \varphi (\infty )} \rangle = \nonumber \\  |x|^{2(-j_1-j_2+j_3+j_4)} |1-x|^{2(-j_1+j_2-j_3+j_4)} |z|^{2j_2} |1-z|^{2j_3} |x-z|^{2(k-j_1-j_2-j_3-j_4)}  \label{una}
\end{eqnarray}
where $j_1=k/2$ and $j_2=j$. Plugging (\ref{una}) into (\ref{otra}) and rewriting the normalization factor $F_k
(k/2,j,j_3,j_4)$ by using
\begin{equation}
\Upsilon _b (Q\mp x) = \pm \Upsilon _b (x) \frac {\Gamma (bx)\Gamma (b^{-1}x)}{\Gamma (\pm bx)\Gamma (\pm b^{-1}x)}
b^{2x(b^{\pm 1}-b)}  \label{shift}
\end{equation}
and 
\begin{equation}
\Upsilon _b (b^{\pm 1}+ x) = \Upsilon _b (x) \frac {\Gamma (b^{\pm 1}x)}{\Gamma (1-b^{\pm 1}x)} b^{\pm 1\mp 2b^{\pm 1}x}   ,  \label{shiftdos}
\end{equation}
we get the following expression
\begin{eqnarray}
\langle \Phi _{j} (0,0) \Phi _{\frac k2} (x,z) \Phi _{j_3} (1,1)  \Phi _{j_4} (\infty ,\infty )  \rangle = c_k  \left( \pi \frac {\Gamma (1-\frac {1}{k-2})}{\Gamma (1+\frac {1}{k-2})}\right)^{1-j-j_3-j_4} \frac { \Gamma (1+\frac {1-2j}{k-2})}{\Gamma
(\frac {2j-1}{k-2})} \times \nonumber \\ 
\times |x|^{2(-\frac k2 -j+j_3+j_4)}  |1-x|^{2(-\frac k2 +j-j_3 +j_4)}  |z|^{2j_2}  |1-z|^{2j_3} |z-x|^{2(\frac k2-j-j_3-j_4)} \times \label{mo} \\
\times \frac {G(1-k/2+j-j_3-j_4) G(j-k/2+j_3-j_4) G(j-k/2-j_3+j_4) G(k/2-j-j_3-
j_4)}{G(-1) G(1-k+2j) G(1-2j_3) G(1-2j_4) } \nonumber
\end{eqnarray}
where the $k$-dependent function $G$ is defined as usual, namely
\begin{equation}
G(x) = \Upsilon ^{-1}_b(-bx) b^{-b^2x^2-(b^2+1)x}  ,   \label{G}
\end{equation}
and $c_k$ is certain $j$-independent factor which is completely determined by (\ref{shift}) and (\ref{G}). Expression (\ref{mo}) is in exact agreement with the result obtained by Maldacena and
Ooguri for the three-point violating winding correlator (cf. Eqs. (5.25), (5.33) and (E.14) of Ref. \cite{MaldacenaOoguri3}). In fact, after Fourier transforming the expression above (with an appropriate regularization realized by a factor $\lim _{x \to \infty}|x|^{2k} $) one gets the scattering amplitude of the three-string process that violates the winding number conservation in $AdS_3$ space (see also \cite{GiribetNunez3}).

It is remarkable that our deduction of (\ref{mo}) does not make use of the
decoupling equation satisfied by the correlators that include a degenerate
state with $j_1 = k/2$. However, this information is implicit in the
derivation above since it is codified in the factor $F_k (k/2,j,j_3,j_4)$.
Furthermore, the fact that the structure constants develop a delta
singularity when one of the vertex involved in the OPE carries momentum
$j_1=k/2$ is manifested in the condition $n=0$ above.

Recently, another relation between Liouville and WZNW theories was used to describe the string scattering amplitudes in $AdS_3$ \cite{RibaultTeschner,Ribault,GiribetNakayama}. In \cite{Ribault}, the three-point function (\ref{mo}) was shown to be directly connected to the Liouville structure constant. It is worth mentioning that the connection of both theories employed here is a different one which, unlike the one in \cite{RibaultTeschner}, is local in the five Liouville insertions on the sphere.

\subsection{Spectral flow and identities between exact solutions}

First of all, let us notice that we can always consider a rescaling of the Liouville cosmological constant $\mu $ (and the WZNW coupling constant
$\lambda $ as well; see \cite{GiveonKutasov})
in such a way that the KPZ factor in (\ref{fz}) can be made to disappear. This corresponds to shifting the zero mode of the field $\varphi $ in an appropriate way.
We adopt such convention here.

\subsubsection*{Identification between states of discrete representations}

Now, let us define $\tilde J _{\mu } = \frac k2 - j_{\mu }$. As it was pointed out in \cite{GiribetSimeone4}, the transformation $j_{\mu }\to \tilde {J} _{\mu }$ corresponds to reflecting the (four) quantum numbers in the Liouville correlation function, namely $\alpha _{\mu }\to Q-\alpha _{\mu }$. Then, according to (\ref{reflex}), we find
\begin{eqnarray}
{\mathcal A}^{WZNW}_{j_1,j_2,j_3,j_4} (x,z) &=& \frac {X_k (j_1,j_2,j_3,j_4|x,z)}{X_k(\tilde J _1,\tilde J _2,\tilde J _3,\tilde J _4|x,z)} \frac {F_k (j_1,j_2,j_3,j_4)}{F_k (\tilde J _1,\tilde J _2,\tilde J _3,\tilde J _4)} \prod _{\nu =1}^{4} R_b (\alpha _{\nu }) \ {\mathcal A}^{WZNW}_{\tilde J _1,\tilde J _2,\tilde J _3,\tilde J _4} (x,z) \nonumber
\end{eqnarray}
with
\begin{eqnarray}
\frac {X_k(j_1,j_2,j_3,j_4|x,z)}{X_k(\tilde J _1,\tilde J _2,\tilde J _3,\tilde J _4|x,z)} &=& |z|^{2(j_1+j_2-k/2)} |1-z|^{2(j_1+j_3-k/2)} |x|^{2(j_4+j_3-j_2-j_1)} \times \nonumber \\ && \times |1-x|^{2(j_4-j_3+j_2-j_1)} |x-z|^{2(k-j_1-j_2-j_3-j_4)}
\end{eqnarray}
Notice that a factor $|x-z|^{2(k-j_1-j_2-j_3-j_4)}$ arises. This implies the existence of a singularity at the point $x=z$. The arising of this singularity in the solutions of the KZ equation was
pointed out in Ref. \cite{MaldacenaOoguri3}. Similar singularities appear
in the solutions studied in \cite{Petkova}, where an expansion in powers of
$(x-z)$ was considered. This was also discussed in \cite{Ponsot} and
\cite{NicholsSanjay}, where the factors developing poles at $x=z$ were
studied in a similar context (cf. equation (42) of Ref. \cite{Ponsot}). Moreover, in \cite{GiribetSimeone4}
logarithmic singularities at $x=z$ for the configuration $k=j_1+j_2+j_3+j_4$
were analyzed by using the same techniques.

As it was remarked in \cite{MaldacenaOoguri3}, ``[t]he presence of the singularity at $z=x$ is very surprising from the point of view of the worldsheet theory since this is a point in the middle of the moduli spaces. [...] The interpretation of this singularity is again associated with instantonic effects''. In Ref. \cite{GiribetNakayama}, this ``instantonic effects'' in the worldsheet theory were studied in relation with Liouville theory as well, tough in a different framework.

On the other hand, the normalization which connects both solutions takes the form
\begin{eqnarray}
 \frac {F_k (j_1,j_2,j_3,j_4)}{F_k (\tilde J _1,\tilde J _2,\tilde J _3,\tilde J _4)} \prod _{\nu =1}^{4} R_b (\alpha _{\nu }) =\tilde c _k \prod _{\mu =1}^{4} \frac {\Gamma (1+\frac {1-2j_{\mu }}{k-2})}{\Gamma (\frac {2j_{\mu }-1}{k-2})}
\end{eqnarray}
where $\tilde c _k$ is a $k$-dependent factor, determined by (\ref{shift}). This presents poles located at $2j_{\mu }=1+(m+1)(k-2)$ for any non-negative integer $m$. Besides, the zeros of this normalization factor arise at $2j_{\mu }=1-m(k-2)$.

\subsubsection*{Formulae: Acting on two states}
Let us notice that making the change $\alpha _1 \to Q-\alpha _1 , \ \alpha _2 \to Q-\alpha _2$, leaving $\alpha _3 $ and $\alpha _4$ unchanged, is equivalent to transforming $j _1 \to k/2 -j _2 $ and $j _2 \to k/2 -j _1 $. In doing this, the factor $X_k (j_1,j_2,j_3,j_4|x,z)/X_k (\tilde {J}_2,\tilde {J}_1,j_3,j_4|x,z)$ stands, and this develops a factor $|x-z|^{2(k-j_1-j_2-j_3-j_4)}$. Indeed, such factor appears every time the transformations of indices $j_{\mu }$ correspond to reflecting the second Liouville vertex $V_{\alpha _1}(z)$. Then, we get
\begin{equation} 
{\mathcal A}^{WZNW}_{j_1,j_2,j_3,j_4} (x,z) = \frac { R_b(\alpha _1)R_b(\alpha _2) F_k (j_1,j_2,j_3,j_4)X_k (j_1,j_2,j_3,j_4|x,z)}{F_k (\tilde {J}_2,\tilde {J}_1,j_3,j_4)X_k (\tilde {J}_2,\tilde {J}_1,j_3,j_4|x,z)}{\mathcal A}^{WZNW}_{\tilde {J}_2,\tilde {J}_1,j_3,j_4} (x,z)
\end{equation}
Actually, this identity motivates the way of computing the violating winding correlator (\ref{mo}), since it precisely involves a transformation $j_1=k/2 \to \tilde {J}_1=0 , \ j_2=j \to \tilde {J}_2=k/2-j$ of its quantum numbers.

\section{Hidden $\Z _2$ symmetry transformations in the four-point function}

\subsection{Liouville reflection and KZ equation}

Now, we can study a different (tough closely related) class of symmetry transformation. Let us define $\hat J _{\mu } = \frac 12 \sum _{\nu =1}^{4}j_{\nu } - j_{\mu +2}$, with $\mu = \{ 1,2,3,4 \}$ and $j_{\mu } = j_{\mu '}$ if $\mu = \mu '$ Mod $4$. The non-diagonal involution $j_{\mu }\to \hat {J}_{\mu }$ is, according to (\ref{mapa}), equivalent to doing $\alpha _3 \to Q-\alpha _3$. This disentangles the symmetry transformation enabling us to understand it as a simple reflection (\ref{reflex}). Consequently, we find the simple relation
\begin{eqnarray}
{\mathcal A}^{WZNW}_{j_1,j_2,j_3,j_4} (x,z) &=& \frac {X_k (j_1,j_2,j_3,j_4|x,z)}{X_k(\hat J _1,\hat J _2,\hat J _3,\hat J _4|x,z)} \frac {F_k (j_1,j_2,j_3,j_4)}{F_k (\hat J _1,\hat J _2,\hat J _3,\hat J _4)} R_b (\alpha _3)  \ {\mathcal A}^{WZNW}_{\hat J _1,\hat J _2,\hat J _3,\hat J _4} (x,z)  \label{andreev}
\end{eqnarray}
with 
\begin{equation}
\frac {X_k(j_1,j_2,j_3,j_4|x,z)}{X_k(\hat J _1,\hat J _2,\hat J _3,\hat J _4|x,z)} = |z|^{\frac {1}{k-2}(j_1-j_2)^2-\frac {1}{k-2}(j_3-j_4)^2} |1-z|^{\frac {1}{k-2}(j_2-j_4)^2-\frac {1}{k-2}(j_1-j_3)^2} |1-x|^{2(j_4-j_3+j_2-j_1)} \nonumber
\end{equation}
In reference \cite{NicholsSanjay}, Nichols y Sanjay proposed that both sides in eq. (\ref{andreev}) ``can, presumably by uniqueness of the solution, be identified [...]; at least up to some overall scale''. The identity above proves such affirmation presenting the precise overall scale, which is found to be
\begin{eqnarray}
\frac {F_k (j_1,j_2,j_3,j_4)}{F_k (\hat J _1,\hat J _2,\hat J _3,\hat J _4)} R_b (\alpha _3) =\hat {c} _k \prod _{\mu =1}^{4}\frac {G (1-2\hat {J} _{\mu })}{G (1-2j _{\mu })}
\end{eqnarray}
where $\hat c _k$ is certain $k$-dependent factor. The factor above develops poles at $\sum _{\nu =1}^{4}j_{\nu } - 2 j_{\mu +2} = (1-m)-n(k-2)$ as well as at $\sum _{\nu =1}^{4}j_{\nu } - 2 j_{\mu +2}= (n+2)+(m+1)(k-2)$, for any ($m,n$) pair of non-negative integers.

Besides, analogous identities are obtained by considering $\alpha _i \to Q-\alpha _i$ with $i\in \{2,3,4\}$, corresponding to 
\begin{eqnarray}
j_1 \to \frac 12 (j_1+j_2+j_3+j_4-2j_i) \ , \ \  j_i \to \frac 12 (j_1+j_2+j_3+j_4-2j_1) \nonumber \\
j_j \to \frac 12 (j_1+j_2+j_3+j_4-2j_k) \ , \ \  j_k \to \frac 12 (j_1+j_2+j_3+j_4-2j_j)
\end{eqnarray}
for any even permutation of the symbol $\{ i,j,k \} $. 

The conciseness of our deduction of (\ref{andreev}) turns out to be surprising. In reference \cite{Andreev} Andreev stated the validity of such an identity and considered that the problem of ``understand[ing] what underlies this mysterious relation'' remained open. Regarding this, it was suggested that ``[m]ay be there is a hidden symmetry in the theory''. Equation (\ref{andreev}) manifestly shows that such hidden symmetry actually corresponds to the Liouville reflection realized by (\ref{fz}).  

\subsubsection*{A working example}

The relation (\ref{andreev}) also enables us to write down the explicit form of certain particular correlators. For instance, a simple computation leads to
\begin{eqnarray}
\langle \Phi _{j_2} (0,0) \Phi _{j_1} (x,z) \Phi _{j_1+j_2+j_4} (1,1) \Phi _{j_4} (\infty ,\infty )\rangle = F_k (j_1,j_2,j_1+j_2+j_4,j_4) R_b (bj_1+Q/2)\times \nonumber \\ 
\times |z|^{-\frac {4}{k-2}j_1 j_2} |1-z|^{-\frac {4}{k-2}j_1 (1-j_1-j_2-j_4)} |1-x|^{-4j_1} \nonumber
\end{eqnarray}
which is worked out in easy way since, as before, it is associated to a Liouville correlator with no insertion of screening charges. Besides, this correlator can be interpreted in terms of  the operator product expansion of exponential operators in the $SL(2,\C)_k/SU(2)$ WZNW model. Certainly, the asymptotic form of the differential functions on $H^+_3$ is governed by exponential functions which, for the configuration above, correspond to the expectation value
\begin{equation}
\langle e^{\sqrt{\frac {2}{k-2}}j_{2}\phi (0)}e^{\sqrt{\frac {2}{k-2}}j_{1}\phi (z)}e^{\sqrt{\frac {2}{k-2}}(1-j_{3})\phi (1)}e^{\sqrt{\frac {2}{k-2}}j_{4}\phi (\infty )} \rangle . \label{veintitres}
\end{equation}
In a stringy theoretical context, this represents a four-string scattering process in $AdS_3$ in which the third
vertex (having a mass $m\sim j_3$ in the large $k$ description) has a particular behavior with respect to the boundary variables ($x_3,\bar {x}_3$) of the space on which the dual CFT is formulated (cf. section 3 of \cite{MaldacenaOoguri3} for a detailed discussion on the large $k$ interpretation of this exponential functions).

\subsection{A $k$-dependent $\Z _2$ symmetry transformation}

Now, let us analyze a combination of the spectral flow and the $\Z _2$ symmetry transformations considered above. More precisely, let us perform the charge $\alpha _1 \to Q- \alpha _1$. In terms of the $SL(2,\R )$ quantum numbers, this corresponds to doing $j_{\mu }\to J _{\mu }= \frac 12 \sum _{\nu =1 } \tilde J _{\nu } - \tilde J _{\mu }$. Hence, the following equality between correlators is found to hold
\begin{eqnarray}
{\mathcal A}^{WZNW}_{j_1,j_2,j_3,j_4} (x,z) &=& \frac {X_k (j_1,j_2,j_3,j_4|x,z)}{X_k(J _1, J _2, J _3,J _4|x,z)} \frac {F_k (j_1,j_2,j_3,j_4)}{F_k ( J _1,J _2,J _3,J _4)} R_b (\alpha _1)  \ {\mathcal A}^{WZNW}_{J _1,J _2,J _3,J _4} (x,z)
\end{eqnarray}
where
\begin{eqnarray}
\frac {X_k(j_1,j_2,j_3,j_4|x,z)}{X_k(J _1,J _2,J _3,J _4|x,z)} = |z|^{\frac {4}{k-2}(J_1 J_2-j_1 j_2)} |1-z|^{\frac {4}{k-2}(J_1 J_3 - j_1 j_3)} |z-x|^{2(k-j_1-j_2-j_3-j_4)}
\end{eqnarray}
Here, the factor $|z-x|^{2(k-j_1-j_2-j_3-j_4)}$ stands again. In terms of the original indices $j_{\mu }$ this transformation reads
\begin{eqnarray}
j_1 \to \frac 12 (k+j_1-j_2-j_3-j_4) \ , \ \  j_2 \to \frac 12 (k-j_1+j_2-j_3-j_4) \nonumber \\
j_3 \to \frac 12 (k-j_1-j_2+j_3-j_4) \ , \ \  j_4 \to \frac 12 (k-j_1-j_2-j_3+j_4) \label{yesta}
\end{eqnarray}
On the other hand, the relative normalization has the form
\begin{equation}
\frac {F_k (j_1,j_2,j_3,j_4)}{F_k ( J _1,J _2,J _3,J _4)} R_b(\alpha _1) =c_k \prod _{\mu =1}^{4} \frac {G (1-2J_{\mu })}{G (1-2j_{\mu })}
\end{equation}
Involution (\ref{yesta}) is a new symmetry of KZ equation, which results uncovered when is interpreted as a simple transformation (\ref{reflex}). This enables us to write down the explicit expression of another special case; namely
\begin{eqnarray}
\langle \Phi _{j_2} (0,0) \Phi _{j_1} (x,z) \Phi _{j_1+k-j_2-j_4} (1,1) \Phi _{j_4} (\infty ,\infty )\rangle = F_k (j_1,j_2,j_1+k-j_2-j_4,j_4) R_b (bj_1+Q/2) \times \nonumber \\ 
\times |z|^{-\frac {4}{k-2}j_1 j_2} |1-z|^{-\frac {4}{k-2}j_1 (1-j_1-k+j_2+j_4)} |z-x|^{-4j_1}
\end{eqnarray}
which, again, can be thought of as the operator product expansion (\ref{veintitres}) of the operators in the $SL(2,\C)_k$ WZNW model. This solution confirms the appearance of the factor $|x-z|^{2(k-j_1-j_2-j_3-j_4)}$ in the expression of several four-point functions \cite{MaldacenaOoguri3}.

\section{Discussion}

Most of what we know about non-compact WZNW model is based on analogies with the Liouville field theory. Moreover, the
Fateev-Zamolodchikov dictionary (\ref{fz}) turned out to be one the principal tools in working out the formal aspects of
this class of conformal models.

Here, we have made use of the relation between correlators in both Liouville and WZNW theories to prove several identities between exact
solutions of the Knizhnik-Zamolodchikov equation. Of course, such identities can, in principle, be proven by explicit
manipulation of the KZ equation, at least up
to a $j$-dependent overall factor \cite{NicholsSanjay}. However, the method of proving them by means of the connection with Liouville CFT turns out to be surprisingly concise and exploits the fact that the structure of the OPE of the
models is encoded in the normalization factor (\ref{F}). The derivation of the three-point function violating the
winding number conservation we presented here turns out to be a good example of this conciseness.

%\newpage

\[
\]

This work was partially supported by Universidad de Buenos Aires. I am grateful to the people of Centro de Estudios
Cient\'{\i}ficos CECS, Valdivia, for the hospitality during my stay, where part of this work was done. I would also
like to express my gratitude to Yu Nakayama for several discussions on related subjects. It is also a pleasure to thank Marco Matone and the organizers of the ``Workshop on Non-Perturbative Gauge Dynamics'', at SISSA, Trieste, 2005.


\begin{thebibliography}{10}

\bibitem{MaldacenaOoguri1} J.M. Maldacena and H. Ooguri, {\it Strings in $AdS_3$ and the $SL(2,\R )$ WZW Model. Part 1: The Spectrum}, J. Math. Phys. {\bf 42} (2001) pp. 2929-2960, [arXiv:hep-th/0001053].

\bibitem{Halpern} K. Bardakci and M.B. Halpern, Phys. Rev. {\bf D3} (1971) 2493. A. Schwimmer and N. Seiberg,  Phys. Lett. {\bf B184} (1987) 191. J.K. Freericks and M.B. Halpern, Ann. of Phys. {\bf 188} (1988) 258; erratum, ibid. {\bf 190} (1989) 212. M.B. Halpern, E. Kiritsis, N.A. Obers and K. Clubok, Physics Reports {\bf 265}, 1 and 2 (1996) 1.


\bibitem{NicholsSanjay} A. Nichols and Sanjay, {\it Logarithmic operators in the $SL(2,\R )$ WZNW model}, Nucl. Phys. {\bf B597} (2001) pp. 633-651, [arXiv:hep-th/0007007].

\bibitem{GiribetSimeone4} G. Giribet and C. Simeone, {\it Liouville theory and logarithmic solutions to Knizhnik-Zamolodchikov equation}, in Press in Int. J. Mod. Phys. {\bf A} (2005), [arXiv:hep-th/0402206].

\bibitem{Nakayama} Yu Nakayama, {\it Liouville Field Theory - A decade after the revolution}, Int.J.Mod.Phys. {\bf A19} (2004) pp. 2771-2930, [arXiv:hep-th/0402009].

\bibitem{KutasovSeiberg} D. Kutasov and N. Seiberg, {\it More Comments on String Theory on $AdS_3$}, JHEP {\bf 9904} (1999) pp. 008, [arXiv:hep-th/9903219].

\bibitem{FateevZamolodchikov} A. Zamolodchikov and V. Fateev, Sov. J. Nucl. Phys. {\bf 43}, 4 (1986) pp. 657.

\bibitem{Andreev} O. Andreev, {\it Operator algebra of the $SL(2)$ conformal field theories}, Phys. Lett. {\bf B363} (1995) pp. 166, [arXiv:hep-th/9504082].

\bibitem{Teschner} J. Teschner, {\it Crossing Symmetry in the $H_3^+$ WZNW model}, Phys. Lett. {\bf B521} (2001) pp.127-132, [arXiv:hep-th/0108121].

\bibitem{Ponsot} B. Ponsot, {\it Monodromy of solutions of the Knizhnik-Zamolodchikov equation: SL(2,C)/SU(2) WZNW model}, Nucl. Phys. {\bf B642} (2002) pp. 114-138, [arXiv:hep-th/0204085].


\bibitem{MaldacenaOoguri3} J.M. Maldacena and H. Ooguri, {\it Strings in $AdS_3$ and the $SL(2,\R )$ WZW Model. Part 3: Correlation Functions}, Phys. Rev. {\bf D65} (2002) 106006, [arXiv:hep-th/0111180].

\bibitem{GiribetNunez3} G. Giribet and C. N\'{u}\~{n}ez, {\it Correlators in $AdS_3$ string theory}, JHEP {\bf 0106} (2001) pp. 010, [arXiv:hep-th/0105200].

\bibitem{RibaultTeschner} S. Ribault and J. Teschner, {\it  $H(3)+$ correlators from Liouville theory}, JHEP {\bf 0506} (2005) pp. 014, [arXiv:hep-th/0502048].

\bibitem{Ribault} S. Ribault, {\it Knizhnik-Zamolodchikov equations and spectral flow in AdS3 string theory}, [arXiv:hep-th/0507114].

\bibitem{GiribetNakayama} G. Giribet and Yu Nakayama, {\it The Stoyanovsky-Ribault-Teschner Map and String Scattering Amplitudes}, [arXiv:hep-th/0505203].

\bibitem{GiveonKutasov} A. Giveon and D. Kutasov, {\it Notes on $AdS_3$}, Nucl. Phys. {\bf B621} (2002) pp. 303-336, [arXiv:hep-th/0106004].

\bibitem{Petkova} P. Furlan, A. Ganchev and V. Petkova, {\it $A_1^{(1)}$ Admissible Representations - Fusion Transformations and Local Correlators}, Nucl. Phys. {\bf B491} (1997) pp. 635-658, [arXiv:hep-th/9608018].

\end{thebibliography}
\end{document}